\documentclass[sigconf,nonacm]{acmart}

\AtBeginDocument{%
  \providecommand\BibTeX{{%
    Bib\TeX}}}

\setcopyright{acmlicensed}
\copyrightyear{2026}
\acmYear{2026}
\acmDOI{XXXXXXX.XXXXXXX}
\acmConference[WWW '26]{The Web Conference 2026}{April 13--17, 2026}{Dubai, UAE}

\acmISBN{978-1-4503-XXXX-X/2026/04}

\usepackage[table]{xcolor}
\usepackage{hyperref}
\usepackage{url}
\usepackage{colortbl}
\usepackage{tabularx}
\usepackage{arydshln} 
\usepackage{float}
\usepackage[linesnumbered,ruled,vlined]{algorithm2e}
\usepackage{adjustbox}
\usepackage{multirow}
\usepackage{booktabs}   
\usepackage{multirow}   
\usepackage{makecell}   
\usepackage{natbib}
\usepackage{amsmath,amsfonts}
\usepackage{graphicx}
\usepackage{textcomp}
\usepackage{pifont}
\usepackage{threeparttable}
\usepackage{comment}
\usepackage{fbox}

\usepackage{amssymb}   
\usepackage{textcomp}  

\newcommand{\qw}[1]{\textcolor{magenta}{#1}}

\newenvironment{packeditemize}{
	\begin{list}{$\bullet$}{
			\setlength{\labelwidth}{4pt}
			\setlength{\itemsep}{0pt}
			\setlength{\leftmargin}{\labelwidth}
			\addtolength{\leftmargin}{\labelsep}
			\setlength{\parindent}{0pt}
			\setlength{\listparindent}{\parindent}
			\setlength{\parsep}{0pt}
			\setlength{\topsep}{1pt}}}{\end{list}}

\usepackage{subcaption} 
\usepackage[font=normalsize, labelfont=bf, labelsep=colon]{caption}  
\def\BibTeX{{\rm B\kern-.05em{\sc i\kern-.025em b}\kern-.08em
    T\kern-.1667em\lower.7ex\hbox{E}\kern-.125emX}}

\usepackage{enumitem}
\setlist[itemize]{noitemsep, topsep=0pt, parsep=0pt, partopsep=0pt,
                  left=0pt, itemsep=0pt, listparindent=\parindent,
                  labelsep=4pt, leftmargin=*}
\renewcommand\footnotetextcopyrightpermission[1]{} 
\begin{document}


\title{Eclipse Attacks on Ethereum’s Peer-to-Peer Network}


\author{Ruisheng Shi$^{1,\star}$, Yuxuan Liang$^{1}$, Zijun Guo$^{1}$, Qin Wang$^{2}$,  \\ Lina Lan$^{1,\star}$,  Chenfeng Wang$^{1}$, Zhuoyi Zheng$^{1}$}
\thanks{Accepted by \textcolor{violet}The ACM Web Conference ({WWW)} 2026. \\ $\star$ denotes corresponding authors.}
\affiliation{
\smallskip
\textit{$^1$Beijing University of Posts and Telecommunications} $|$ \textit{$^2$ UNSW Sydney} \country{}
}

\renewcommand{\shortauthors}{}
\renewcommand{\shorttitle}{}

\begin{abstract}
Eclipse attacks isolate blockchain nodes by monopolizing their peer-to-peer connections. The attacks were extensively studied in Bitcoin (SP'15, SP'20, CCS'21, SP'23) and Monero (NDSS'25), but their practicality against Ethereum nodes remains underexplored, particularly in the post-Merge settings. 

We present the first end-to-end implementation of an eclipse attack targeting Ethereum (2.0 version) execution-layer nodes. Our attack exploits the bootstrapping and peer management logic of Ethereum to fully isolate a node upon restart. We introduce a multi-stage strategy that majorly includes (i) poisoning the node's discovery table via unsolicited messages, (ii) infiltrating Ethereum’s DNS-based peerlist by identifying and manipulating the official DNS crawler, and (iii) hijacking idle incoming connection slots across the network to block benign connections. Our DNS list poisoning is the first in the cryptocurrency context and requires only 28 IP addresses over 100 days. Slots hijacking raises outgoing redirection success from 45\% to 95\%.
We validate our approach through controlled experiments on Ethereum's Sepolia testnet and broad measurements on the mainnet. Our findings demonstrate that over 80\% of public nodes do not leave sufficient idle capacity for effective slots occupation, highlighting the feasibility and severity of the threat. We further propose concrete countermeasures and responsibly disclosed all findings to Ethereum's security team.
\end{abstract}


\maketitle

\section{Introduction}\label{sec:1}

Ethereum has become one of the most successful cryptocurrency platforms. On Sep. 15, 2022, Ethereum completed its long-anticipated merge~\cite{website:Merge}, transitioning from Proof-of-Work (PoW) to Proof-of-Stake (PoS). This merge unified the original Ethereum mainnet (execution layer) and the Beacon Chain (consensus layer), reducing energy consumption by 99.95\% by eliminating the need for energy-intensive mining. This transition introduced new assumptions and behavioural changes in the peer-to-peer (P2P) network layer \cite{website:node_client}:

\begin{packeditemize}
\item \textbf{Long-lived validator nodes.} PoS validators maintain persistent online presence, resulting in more stable peer sets.
\item \textbf{Static bootstrapping and peer discovery.} Validators often reuse static peer configurations or depend heavily on DNS-based discovery mechanisms. 
\item \textbf{Client usage patterns.} A small number of client pairings (e.g., Prysm+Geth) dominate validator deployments.
\item \textbf{Exposure of execution clients.} Although block finality is determined by the consensus layer, the execution client handles transaction gossip and mempool operations.
\end{packeditemize}

These shifts call for a renewed assessment of classical network-layer threats within Ethereum’s PoS architecture. For instance, the persistent online presence of validators results in stable peer sets, making them ideal targets for long-term network manipulation. The reliance on static bootstrapping and DNS-based discovery significantly enhances the success rate of peer poisoning strategies.

In this paper, we revisit and refine the classic eclipse attack under these updated assumptions. We propose a practical, multi-stage attack that leverages previously discussed vectors (e.g., discovery table poisoning, DNS list infiltration) to strategically occupy both incoming and outgoing connections.

We present our progress in a step-by-step manner.

\begin{packeditemize}
\item[$\triangleright$] \textbf{We systematically analyze the P2P networking behavior of post-Merge Ethereum execution-layer nodes. }
\end{packeditemize}

We deeply analyze how post-Merge Ethereum nodes discover peers, manage connections, and maintain the network. Our study covers both client architecture and address formats, particularly the transition from legacy enode to the more extensible ENR format. We delve into the structure and management of both short-term (discovery table) and long-term (DB) peer records, highlighting mechanisms such as Discv4 message handling, dynamic K-bucket updates, IP-based admission limits, and initialization routines that protect against adversarial influence.

In addition, we explore the full connection management lifecycle, including the establishment and limitations of incoming and outgoing connections, rate-limiting policies, and the fallback to DNS-based discovery. We further examine the role of the \texttt{devp2p} crawler in maintaining Ethereum’s DNS peerlist, detailing its scoring system, update frequency, and pruning behavior. These insights reveal points of vulnerability within Ethereum’s peer discovery and connection protocol stack, which inform our attack design and defense proposals.

\begin{packeditemize}
\item[$\triangleright$]  \textbf{{We propose the first practical eclipse attack against the Ethereum’s post-Merge P2P network.}}
\end{packeditemize}

Eclipse attacks isolate a target node by controlling all its P2P connections~\cite{apostolaki2017hijacking,tran2020stealthier,baek2023sustainability,saad2023three}. Once isolated, an attacker can feed the victim false information, undermining blockchain consistency and enabling various attacks—such as deanonymization~\cite{biryukov2014deanonymisation,fanti2017deanonymization,biryukov2019linkability}, selfish mining~\cite{feng2019selfish,ritz2018impact,kang2021understanding,tiwari2018distributed}, double spending, or smart contract manipulation~\cite{chen2020understanding}. These attacks reduce network resilience and can support larger threats like 51\% attacks.

Research on Ethereum’s network layer has explored its security~\cite{chen2020survey}, topology~\cite{li2025place,kim2018measuring}, and P2P behavior~\cite{heimbach2024validators}, and attacks such as partitioning and eclipsing~\cite{marcus2018low,dahlke2018low,henningsen2019eclipsing,wust2016ethereum,heo2023partitioning,saad2023three}, topology mapping~\cite{gencer2018decentralization,kim2018measuring,maeng2020analysis,masoud2024measurement,lee2020measurements,kiffer2021gossip,eisenbarth2022monitoring,gao2019topology,li2021toposhot,zhao2024dethna,li2025place,silva2020impact}, and network-layer deanonymization~\cite{gao2021deanonymization,klusman2023deanonymisation,beres2021profiling,lin2024denseflow,heimbach2024validators,wang2024rpc,zhou2022behavior,chan2017ethereum}. However, most eclipse attack studies focus on Bitcoin~\cite{heilman2015eclipse,tran2021routing,yves2018total} or Monero~\cite{shi2025eclipse}, and existing Ethereum work largely predates the Merge~\cite{marcus2018low,dahlke2018low,henningsen2019eclipsing}.

\begin{packeditemize}
\item[$\triangleright$] \textbf{{We have new designs for implementing the attack.}}
\end{packeditemize}

Ethereum’s current network stack introduces new challenges for eclipse attacks. Recent Geth versions preload benign peers into the discovery table, blocking early attacker entries. Since v1.9.11, Ethereum also uses DNS-based peer discovery, which can bypass poisoned tables. Finally, each node only accepts 34 incoming connections, making slot preemption essential.

We design a multi-stage attack. We first fill the target’s peerlist (discovery table and DNS), then hijack incoming slots network-wide. When the target restarts, it is forced to connect only to attacker-controlled nodes so as to complete the eclipse. 

Here, we highlight our new designs.

\begin{packeditemize}
\item[$\triangleright$]   \textbf{{We evaluate attacks on both Sepolia testnet and mainnet.}}
\end{packeditemize}

We implement and evaluate our attack pipeline on Sepolia (testnet), validating the feasibility and success rate of each attack stage under realistic network conditions. To assess applicability, we also analyze Ethereum mainnet nodes and measure network-wide slot availability, connection patterns, and peer behavior.

We conduct a step-by-step experimental analysis of each attack component. For instance, in evaluating the incoming slot hijacking strategy, we scan over 2000 public Ethereum nodes and observe that more than \textbf{80\%} of them have at most \textbf{10} available incoming slots, confirming the feasibility of large-scale slots occupation. Our DNS poisoning strategy is tested over a 3-month period, during which we successfully infiltrate the DNS list with attacker-controlled IPs. Experiments are carefully documented, and key metrics (e.g., filling rates, IP resource usage, timing thresholds) are quantified in \S\ref{sec:4}.

\begin{packeditemize}
\item[$\triangleright$] \textbf{We take further responsible actions.}
\end{packeditemize}

To promote the security of the Ethereum ecosystem, we have taken responsible steps throughout the course of this work.

  \begin{center}
    \fbox{%
    \begin{minipage}{0.9\linewidth}
      All our identified vulnerabilities and attack workflow were responsibly disclosed to Ethereum official team.
    \end{minipage}
   }
   \vspace{5pt}
  \end{center}
  
Alongside these disclosures, we proposed concrete mitigation strategies, including immediate implementation-level patches as well as longer-term protocol design improvements.


Last, we summarized our contributions in short:

\begin{itemize}
    \item full exploration on Ethereum’s P2P network (\S\ref{sec:2});
    \item a new practical, multi-stage eclipse attack  (\S\ref{sec:3});
    \item evaluation on testnet and mainnet (\S\ref{sec:4});
    \item countermeasure suggestions (\S\ref{sec-mitigation}).
\end{itemize}

\section{Dissecting Ethereum P2P Network}
\label{sec:2}

We analyze the network-layer behavior of Ethereum nodes based on \texttt{Geth v1.14.3}~\cite{website:geth_version} (released May 9, 2024), focusing on: (i) node roles/types; (ii) peerlist management; (iii) connection handling and (iv) DNS peerlist management.

\subsection{Ethereum Nodes}
\label{sec:2.1}

An Ethereum node consists of two coordinated clients: the \textit{execution}\footnote{This paper focuses exclusively on the execution client. Unless otherwise stated, references to an “Ethereum node” refer to the execution client.} client and the \textit{consensus} client~\cite{website:node}. The execution client handles transaction propagation, while the consensus client is responsible for block propagation and participation in the PoS consensus protocol. Each client connects to their P2P network.

\subsection{Peerlist Management}
\label{sec:2.2}

To support efficient node discovery and connection establishment, Ethereum employs a distributed and dynamically updated peerlist management mechanism.

\smallskip
\noindent\textbf{Short-\&long-term databases.}\label{sec:2.2.1}
The local node maintains a long-term node storage database (referred to as \textit{DB}) with no size limitations. This database persists even after the local node is restarted, retaining all stored nodes.
Concurrently, each time the local node starts up, it creates a short-term node database in the memory called the \textit{discovery table}, used for node discovery and connection management within the Ethereum network. Unlike the long-term database, this short-term one is cleared entirely upon each restart of the local node.

\smallskip
\noindent\textbf{Discv4 messages.}\label{sec:2.2.2}
Ethereum's node discovery is implemented via the Discv4 protocol~\cite{website:discv4}, which operates over UDP. 
The protocol defines six types of message packets, including \texttt{Ping}/\texttt{Pong} for liveness detection and node table updates, 
\texttt{FindNode}/\texttt{Neighbors} for locating the closest nodes in the ID space, 
and \texttt{ENRRequest}/\texttt{ENRResponse} for exchanging node records. 
Together, these messages enable nodes to test connectivity, discover new peers, and maintain an up-to-date view of the network topology.


\smallskip
\noindent\textbf{Management mechanism of \textit{DB}.}\label{sec:2.2.3}
The long-term database (i.e., \textit{DB}) maintains metadata for each discovered node. 
Two main operations are performed:

\begin{packeditemize}
    \item \textit{Node addition:} A node is added to the database if it has remained in the discovery table for over 5 minutes and has passed at least one liveness check. This check is performed in a recurring cycle every 30 seconds.
    
    \item \textit{Node deletion:} Nodes that have not responded to \texttt{Ping} messages within the last 24 hours (as indicated by the \texttt{lastpong} timestamp) are removed from the database. This cleanup process is triggered every hour.
\end{packeditemize}

\smallskip
\noindent\textbf{Data structure of discovery table.}\label{sec:2.2.4}
The discovery table comprises 17 K-buckets, each structured according to the Kademlia protocol to manage and organize peer information in a P2P network. Nodes are grouped based on their XOR distance from the local node and are ranked by activity. Since every bucket stores up to 16 active nodes, the table can accommodate up to $272$ ($=17 \times 16$) nodes.



To prevent overrepresentation of nodes from the same subnet, Ethereum enforces two IP-level restrictions:
\begin{packeditemize}
    \item \texttt{bucketIPLimit:} At most 2 nodes from the same /24 subnet may reside in any single bucket.
    \item \texttt{tableIPLimit:} Across the entire discovery table, a maximum of 10 nodes from the same /24 subnet is allowed.
\end{packeditemize}
 
\smallskip
\noindent\textbf{Initialization of discovery table.}\label{sec:2.2.5}
After the local node starts, the table is initialized through the following steps:

\begin{packeditemize}
    \item The local node first loads 30 random nodes from the database and adds 4 bootnodes to form 34 seed nodes (via \texttt{loadSeedNodes}). “Random” means generating an ID and selecting the first node in the database with a greater or equal lexicographic ID.
    \item  It then performs a self-lookup and a random lookup to populate the table with benign nodes (via \texttt{doRefresh}, \S\ref{sec:2.2.6}), reducing the risk of attacker-controlled entries.
\end{packeditemize}

Once initialization is complete, the discovery table enters a dynamic update phase, during which it continuously refreshes its entries based on network activity and liveness checks.

\smallskip
\noindent\textbf{Update of discovery table.}\label{sec:2.2.6}
The discovery table is updated via node insertion, deletion, and replacement.

\begin{packeditemize}
    \item \textit{Node filling.} The discovery table is populated through both active and passive mechanisms. In the active case, the node periodically runs the \texttt{doRefresh} function (every 15–30 minutes) to insert seed nodes (\texttt{loadSeedNodes}, see \S\ref{sec:2.2.5}) and perform three random lookups (\texttt{lookupRandom}) to refresh bucket entries. Each lookup initiates a recursive \texttt{FindNode} query based on a randomly generated target ID, collecting \texttt{Neighbors} responses to progressively discover closer nodes, which are then stored in the table. In the passive case, the node adds peers that send it \texttt{Ping} messages during normal operation (lookup recursive search refers to Appendix~\ref{appendix:lookup}).
    

    \item \textit{Node deletion and replacement.} The \texttt{doRevalidate} function runs every 0–10 seconds, randomly selecting a bucket to verify liveness. It pings the latest seen node; if responsive, the node is promoted within the bucket and its ENR is updated if needed. If unresponsive, it is replaced by a node from the replacement list.

\end{packeditemize}

\subsection{Connections Management}
\label{sec:2.3}

By default, the upper bound of connections for an Ethereum node is 50, with a maximum of 16 outgoing and 34 incoming connections. 
 
\smallskip
\noindent\textbf{Incoming connections establishment.}\label{sec:2.3.1}
An Ethereum node allows up to 34 incoming connections. Since incoming connections are passively established, we discuss the reasons why such connection attempts may fail in Appendix~\ref{appendix:incom_fail}.



\smallskip
\noindent\textbf{Outgoing connections establishment.}\label{sec:2.3.2}
A node maintains up to 16 outgoing connections. After startup, once the discovery table is initialized, the node begins establishing outgoing connections via a dial management loop, which continuously attempts to connect to peers if slots are available. Static nodes (manually configured peers) are prioritized but are not considered here as they need to be added manually by users. 
Peers are selected with equal probability from two sources: lookup discovery and DNS discovery. 
See more details of establishing outgoing connections in Appendix~\ref{appendix:select_outcon}.

\smallskip
\noindent\textbf{Connections dropping.}\label{sec:2.3.3}
Dropped connections in the network are primarily caused by network-layer issues, such as TCP connection timeouts or peer nodes going offline.

\subsection{Devp2p Tool Management}\label{sec:2.4}

Ethereum uses the \texttt{devp2p} command-line tool, embedded in the Geth client, to maintain the official DNS peerlist. Nodes listed here only support the UDP-based discovery protocol and are crawled periodically by the \texttt{devp2p} tool.

\smallskip
\noindent\textbf{Crawler.}\label{sec:2.4.1} Each \texttt{devp2p} node concurrently performs: (i) updates to nodes listed in the input JSON file, and (ii) iterative peer discovery from seed nodes by issuing \texttt{FindNode} queries to known peers.

\smallskip
\noindent\textbf{Scoring policy.}\label{sec:2.4.2} When a \texttt{devp2p} crawler updates a node’s ENR via a successful \texttt{ENRResponse}, the node gains +1 point. Crawlers throttle ENR updates to one every 10 minutes. If a node fails to respond, its score is halved. Once the score drops below 0, the node is evicted from DNS list. Specifically, the devp2p tool awards points to crawled nodes in two scenarios:

\begin{packeditemize}
\item \textit{Startup update bonus.} When a crawl begins, the devp2p node updates all nodes listed in the DNS peerlist. Any node that is found to be online receives +1 point.
\item \textit{Node discovery bonus.} During the crawl, the devp2p node selects peers from its discovery table and sends them FindNode requests. Any node returned in the Neighbors response is also awarded +1 point.
\end{packeditemize}

\noindent\textbf{Crawling schedule.}\label{sec:2.4.3} Ethereum launches a \texttt{devp2p} crawler every 5 hours and 40 minutes. Each crawl lasts 30 minutes, repeating 4–5 times daily. Nodes from all crawls are merged into an aggregate list (\texttt{all.json}), from which the top-N nodes (based on score) are selected per network -- 3000 for mainnet, 250 for testnet.

\section{Our Eclipse Attack}
\label{sec:3}

We target Ethereum execution layer nodes with public IP addresses, occupying incoming and outgoing connections of the target node. 

\subsection{Overview of The Attack}\label{sec:3.1}
\begin{figure*}[!t]
\centering
\includegraphics[width=0.99\linewidth]{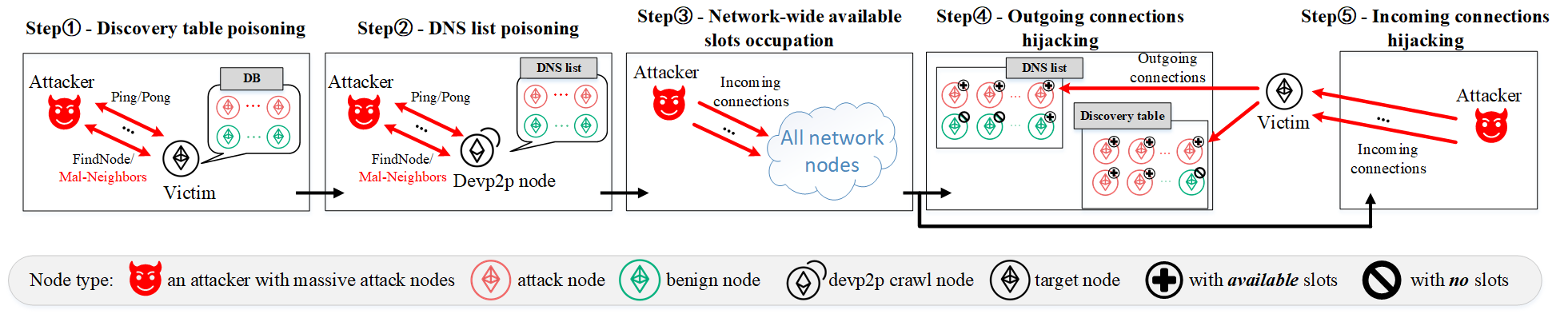} 
\caption{\textbf{Our Eclipse Attack Workflow} ({\small Steps \ding{172}–\ding{174} operate in parallel in practice, we illustrate them sequentially for clarity.)}} 
\label{fig:workflow}
\end{figure*}

Our attack proceeds in five steps (illustrated in Fig.~\ref{fig:workflow}). We present details for each step in the following sections.

\begin{packeditemize}
\item[\ding{172}] \textbf{Discovery table poisoning by \textit{DB} pre-filling.} Before the target node launches, an attacker pre-fills its local database (\textit{DB}) to increase the probability of malicious entries being loaded into the table upon startup.

\item[\ding{173}] \textbf{DNS list poisoning.} An attacker continuously injects malicious ENRs into Ethereum’s DNS-based discovery list prior to target node's restart, polluting a secondary peer source.

\item[\ding{174}] \textbf{Network-wide available slots occupation.} The attacker preemptively fills available incoming connection slots across the network, reducing the number of reachable honest peers when the target node attempts to connect after reboot.

\item[\ding{175}] \textbf{Outgoing connections hijacking.} Upon the target node’s restart, the attacker activates numerous malicious nodes to send \texttt{Ping} messages, filling the discovery table. With most honest nodes’ incoming slots already occupied, the target node is likely to establish outgoing connections with attacker-controlled peers.

\item[\ding{176}] \textbf{Incoming connections hijacking.} Simultaneously, the attacker floods the target node with incoming connection requests,  exhausting incoming connection limits and completing the eclipse.
\end{packeditemize}

\subsection{(\textbf{Step-\ding{172}}) Discovery Table Poisoning\label{sec:3.2} 
}
We begin our attack by pre-filling the target node’s persistent database (i.e., \textit{DB}) to poison its discovery table upon restart. Normally, when a node boots up, it initializes its discovery table with a small set of seed nodes—some randomly selected from \textit{DB}, and others hardcoded as bootnodes. As a result, even before the node starts processing unsolicited \texttt{Ping} messages, part of its discovery table is already occupied by these entries. This limits the attacker’s ability to fully control the table post-restart.

To overcome this limitation, we launch a two-stage attack. First, we inject attacker-controlled nodes into the target’s \textit{DB} before it restarts. Then, upon the next boot, these nodes are likely selected as seed nodes, effectively giving the attacker a foothold in the discovery table from the outset.

To achieve DB poisoning, we send \texttt{Ping} messages from attacker nodes to the target node, leveraging the passive discovery behavior (\S\ref{sec:2.2.6}), which causes the target to insert those nodes into its discovery table. Then, we rely on the node addition logic of the persistent database (\S\ref{sec:2.2.3}), which periodically stores discovery table entries that have passed liveness checks for more than 5 minutes into \textit{DB}.

To maximize poisoning coverage, we perform the attack in multiple rounds, described below:


\vspace{-0.17in}
\begin{center}
\resizebox{\linewidth}{!}{
\fbox{%
\begin{minipage}{0.999\linewidth}
\begin{itemize}
\item We detected the restart of the target node.
\item We immediately send \texttt{Ping} messages from the first round of attacker nodes, which gradually populate both the discovery table and \textit{DB}.
\item Once the first-round nodes are stored in \textit{DB}, we terminate them and launch a second round of attacker nodes. As the first-round nodes go offline, the second-round nodes replace them in the discovery table and eventually in \textit{DB}.
\item We repeat this replacement process across multiple rounds until \textit{DB} is sufficiently filled with attacker-controlled ENRs.
\item Finally, we detect the restart of the target node for a \underline{second} time, triggering a discovery table initialization phase dominated by attacker entries drawn from the poisoned database \textit{DB}.
\end{itemize}
\end{minipage}
}
}
\end{center}

This secondary restart is essential. It ensures that the attack nodes from the poisoned \textit{DB} are loaded early into the discovery table, improving the attacker’s success rate. Although the first restart is needed to initiate the initial discovery table state, subsequent rounds do not require restarts, as discovery table entries can still be updated efficiently while the node is online.

During the target node's second restart, we launch a discovery table poisoning attack concurrently. Specifically, 272 attacking nodes (matching the 272 slots in the discovery table) send \texttt{Ping} messages to the target node to maximize attacker's occupancy of the discovery table. To further improve the occupancy, we modify the behavior of the attacking nodes: upon receiving \texttt{FindNode} messages from the target node, they respond with \texttt{Neighbors} messages that include attacking nodes rather than benign nodes.

\subsection{(\textbf{Step-\ding{173}}) DNS List Poisoning}\label{sec:3.3}

Next, we pre-fill the Ethereum DNS discovery list with attacker-controlled nodes, such that the target node is likely to select our malicious nodes when establishing outgoing connections. Our attack proceeds in two main phases.

We first inject malicious nodes into the master DNS list. We launch a large number of attacker nodes and keep them stably online. Empirically, a mainnet node is likely to be included in the DNS list within 1–2 days, while a Sepolia testnet node takes about 10 days. To improve visibility, we configure nodes with low ENR node IDs (e.g., prefixed with leading zeros) so that they are ranked at the top of crawler input list.

After our nodes are included, we repeatedly trigger score increments to improve their ranking:


\vspace{-0.1in}
\begin{center}
\resizebox{\linewidth}{!}{
\fbox{%
\begin{minipage}{0.999\linewidth}
\begin{itemize}
\item Identify the official \texttt{devp2p} crawler node;
        \item Pre-generate node IDs prefixed with zeros and start nodes to appear first in the crawler’s input file;
        \item Monitor incoming \texttt{ENRRequest} messages to infer the crawler’s launch time;
        \item Activate multiple malicious nodes to \texttt{Ping} the crawler, fill its discovery table, and respond to \texttt{FindNode} queries with neighbors lists containing our own malicious nodes;
        \item Respond to any \texttt{ENRRequest} from the crawler to gain score increments.
\end{itemize}
\end{minipage}
}
}
\end{center}

Once a sufficient number of our nodes reach high scores, they are selected into the top-$N$ official DNS list used by Geth clients at startup, thereby increasing the probability that the target node connects to malicious peers.





\subsection{(\textbf{Step-\ding{174}}) Network-wide Slots Occupation}\label{sec:3.4}
We preemptively occupy the available incoming connection slots of as many benign nodes as possible. This blocks the target node from establishing connections with honest peers; when the target attempts to connect, it is forced to rely on attacker-controlled nodes instead—enabling an eclipse attack.

\subsection{(\textbf{Step-\ding{175}}) Outgoing Connections Hijacking}\label{sec:3.5}
After the target node restarts, all of its existing connections are dropped. We take advantage of this reset: since Ethereum lacks an active connection eviction policy, we activate a swarm of attacker nodes that rapidly send \texttt{Ping} messages to the target. This increases the chances that the target will select these attacker nodes from its discovery table or DNS list for outgoing connections.

\subsection{(\textbf{Step-\ding{176}}) Incoming Connections Hijacking}\label{sec:3.6}
Upon the target node restarts, its incoming connection slots are unoccupied. We launch a large number of malicious nodes that aggressively initiate connection requests to the target, aiming to fill all of its incoming connection slots before benign peers connect.

\section{Evaluation of Our Attack}
\label{sec:4}

We implemented our eclipse attack with details deferred to Appendix~\ref{appendix:pseudo_code_of_the_eclipse_attack}.
Then, we conducted experimental evaluation towards each attack strategy in \S\ref{sec:3}. 

\subsection{Evaluating \textbf{Step-\ding{172}} (Discovery Table)}\label{sec:4.1}

We first investigated the attack targeting Ethereum’s node discovery mechanism. In this strategy, an attacker preemptively sends a large number of crafted node records to potential victim nodes, aiming to insert and maintain attacker-controlled entries in their discovery tables. This manipulation increases the likelihood of the attacker being selected during peer connections, which is the foundation for our eclipse attack. All experiments in this part were conducted on Sepolia testnet.

We begin by measuring the growth scale and storage characteristics of node database to assess the resource cost and practicality of such an attack. Next, we design and validate a feasible filling strategy that complies with the protocol while continuously pushing malicious records to the target. Finally, we conduct controlled experiments to evaluate the effectiveness of the attack, focusing on the proportion of attacker nodes in the victim's discovery table.

Our experimental setup involves deploying attacker and target nodes on two public servers and we disabled IP-based admission limits on the targets. Attack nodes were instantiated to cover slots in the target discovery table and repeatedly sent unsolicited \texttt{Ping} messages to maximize occupancy and evict existing entries. Upon receiving a target's \texttt{FindNode}, attackers should reply with \texttt{Neighbors} containing attacker entries.

\begin{figure*}[htbp]
\centering
\begin{subfigure}[t]{0.32\textwidth}
    \includegraphics[width=\linewidth,height=3cm]{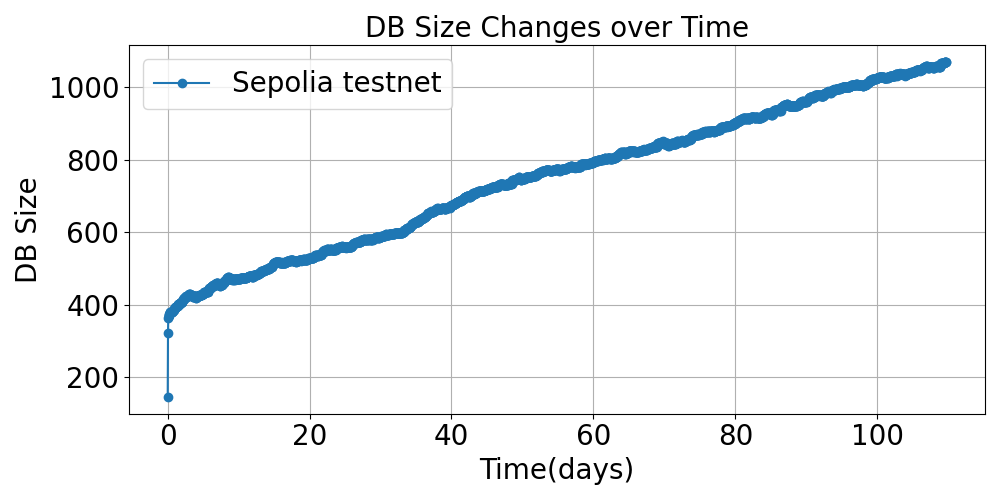}
    \caption{DB Size Changes over Time}
    \label{fig:4}
\end{subfigure}
\hfill
\begin{subfigure}[t]{0.32\textwidth}
    \includegraphics[width=\linewidth,height=3cm]{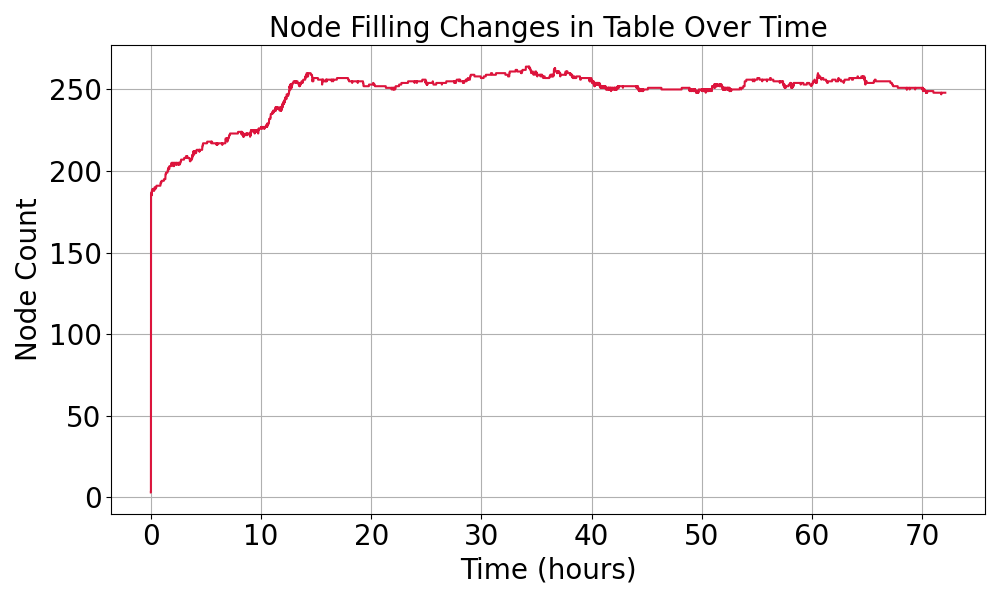}
    \caption{Node Filling Changes over Time}
    \label{fig:node_fill}
\end{subfigure}
\hfill
\begin{subfigure}[t]{0.32\textwidth}
    \includegraphics[width=\linewidth,height=3cm]{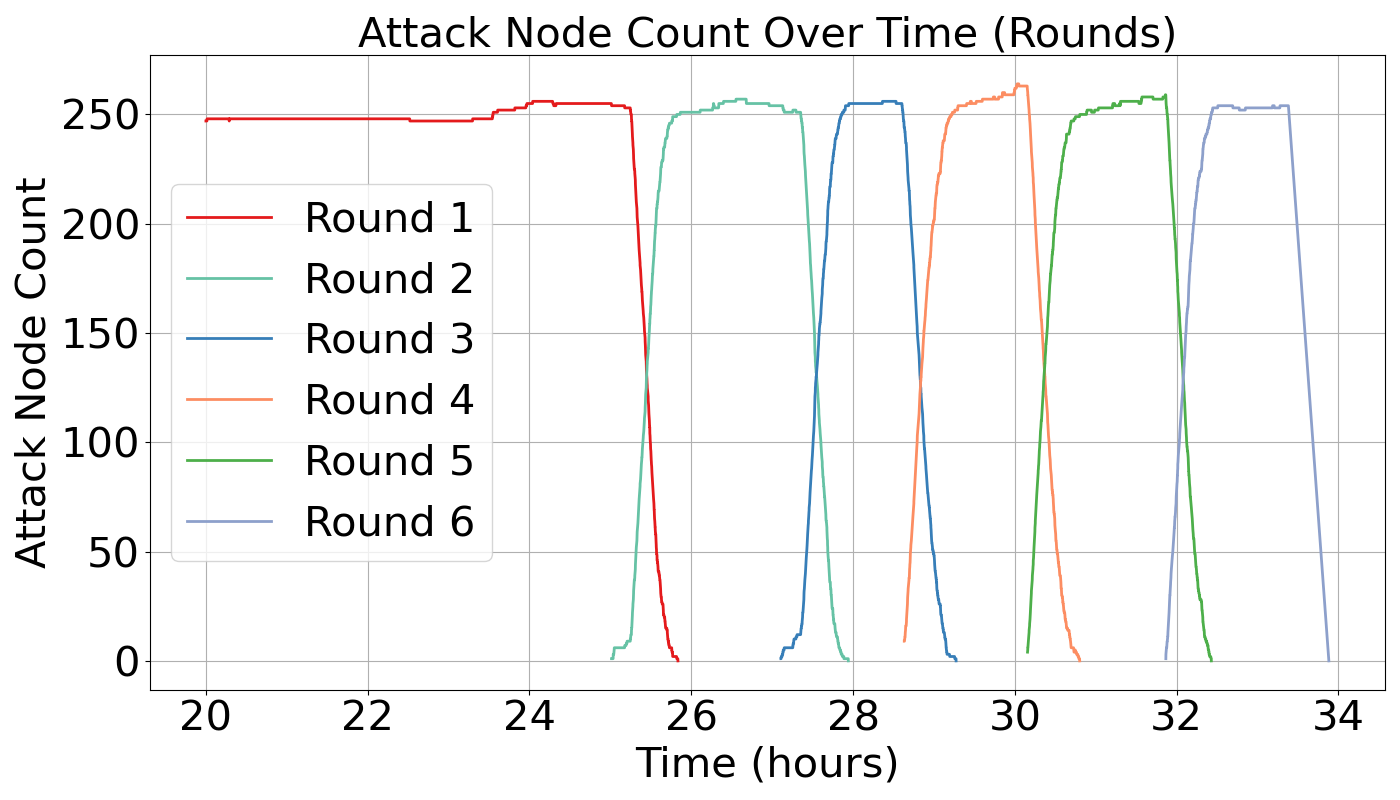}
    \caption{Attack Node Count over Time}
    \label{fig:55}
\end{subfigure}
\caption{Evaluations on Sepolia Testnet}
\label{fig:sepolia_evolution}
\end{figure*}

\smallskip
\noindent\textbf{Measurement of DB growth.}\label{sec:4.1.1}
Since the number of benign nodes in the target’s DB directly affects the success of DB stuffing, we deployed a Sepolia full node on a public network server and monitored its DB growth from an empty state. As shown in Fig.~\ref{fig:4}, the DB grows at an average rate of 200 nodes per month. Based on Ethereum’s official statistics on May 5, 2025, the total number of Sepolia nodes is estimated to be around 2,000. Thus, the maximum realistic DB size on Sepolia is expected to reach 2,000 entries.

\smallskip
\noindent\textbf{Feasibility of poisoning.}\label{sec:4.1.2}
We have two steps. \textit{Initial filling}: we evaluate whether attacker-controlled nodes can be successfully inserted into an initially empty db.
\textit{Repeated insertions}: we evaluate the feasibility of sustaining and expanding DB presence across multiple rounds of poisoning, even after benign entries are added. 


Upon the target's restart, the attacker launched 272 malicious nodes that continuously \texttt{Ping} the target to populate its discovery table; we measured the filling rate over time. The first round steadily increased attacker-controlled entries to ~95\% occupancy within 24 hours (Fig.~\ref{fig:node_fill}, Table~\ref{tab:234}). After terminating that batch, we repeated the procedure with new batches every 2 hours for a total of six rounds. Each new batch was able to refill the table—typically to nearly 100\%—within about 1 hour of the previous batch stopping, demonstrating the effectiveness and repeatability of the multi-round filling strategy (Fig.~\ref{fig:55}, Table~\ref{tab:234}).

\begin{table}[htbp]
\centering
\caption{Node Filling Rates for Discovery Table}
\label{tab:234}
\resizebox{0.85\linewidth}{!}{%
\begin{tabular}{ccccccc}
\toprule
\multicolumn{1}{c}{Round} & \multicolumn{6}{c}{\textbf{Experiment times}} \\
\midrule
& \textbf{1} & \textbf{2} & \textbf{3} & \textbf{4} & \textbf{5} & \textbf{6} \\ 
\cmidrule{2-7}
\multirow{2}{*}{\textbf{1st}} & \cellcolor{gray!10}262/272 & \cellcolor{gray!10}249/272 & \cellcolor{gray!10}255/272 & \cellcolor{gray!10}265/272 & \cellcolor{gray!10}257/272 & \cellcolor{gray!10}260/272 \\ 
& \cellcolor{gray!10}(96.32\%) & \cellcolor{gray!10}(91.54\%) & \cellcolor{gray!10}(93.75\%) & \cellcolor{gray!10}(97.43\%) & \cellcolor{gray!10}(94.49\%) & \cellcolor{gray!10}(95.59\%) \\
\midrule
\multirow{2}{*}{\textbf{2nd}} & \cellcolor{gray!10}256/272 & \cellcolor{gray!10}257/272 & \cellcolor{gray!10}256/272 & \cellcolor{gray!10}264/272 & \cellcolor{gray!10}259/272 & \cellcolor{gray!10}254/272 \\ 
& \cellcolor{gray!10}(94.12\%) & \cellcolor{gray!10}(94.49\%) & \cellcolor{gray!10}(94.12\%) & \cellcolor{gray!10}(97.06\%) & \cellcolor{gray!10}(95.22\%) & \cellcolor{gray!10}(93.38\%) \\
\bottomrule
\end{tabular}
}
\vspace{-0.15in}
\end{table}



\smallskip
\noindent\textbf{Discovery table poisoning.}\label{sec:4.1.3}
From \S\ref{sec:2.2.4}, it is known that the division of buckets is based on the number of leading zero bits of the XOR distance of node IDs. The probability \( P_k \) that a randomly generated ID falls into a certain bucket \( k \) is:
\(
P_k = \frac{1}{2^{16 - k}} - \frac{1}{2^{16 - (k - 1)}} = \frac{1}{2^{17 - k}}
\).


The probability that the target node selects peers from the last $N$ buckets when establishing outgoing connections is summarized in Table~\ref{tab:discovery_filling}. Accordingly, we evaluate the discovery table filling rate based on the occupancy ratio of the last five buckets, as these have the highest likelihood of being used for outgoing connections.



We adjust the number of nodes filled in each round. After experimentation, we observe that the seed nodes selected \textit{randomly} from the DB are typically placed in the last 8 buckets of the discovery table. As a result, subsequent rounds of node filling focus only on these 8 buckets, with each round inserting 128 ($=8×16$) nodes.

 
 
Ethereum nodes establish outgoing connections by randomly generating target node IDs and selecting the closest matching entries from discovery table. Therefore, increasing the occupancy of higher-indexed buckets (i.e., those farther from local node IDs) directly improves the likelihood that outgoing connections are hijacked by attacker-controlled nodes.

We conduct experiments with two initial DB configurations (257 and 517 benign seed nodes) and observe consistent results, summarized in Table~\ref{tab:discovery_filling}. Below, we describe the impact of DB filling attacks using the 257-node case.

Without any DB filling, the last 2 buckets of discovery table remain empty (0\% fill rate), and the last 5 buckets are only 22\% filled. As DB filling rate increases from 33\% to 66\%:

\begin{packeditemize}
    \item Without altering attacker node behavior, the last two buckets’ fill rate increases from 37\% to 70\%, and the last five buckets from 28\% to 44\%.

    \item When attacker node behavior is adjusted (e.g., ID tuning), the fill rate of the last two buckets increases from 36\% to 68\%, and the last five buckets from 22\% to 76\%.
\end{packeditemize}

These results show that: (i) Standard DB filling effectively improves the occupancy of the last two buckets. (ii) DB filling with attacker behavior adaptation significantly boosts the occupancy of the last five buckets, thereby enhancing the success of outgoing connection hijacking.

\begin{table*}[!t]
\centering
\caption{\textbf{Poisoning Discovery Table by DB Pre-filling} (\textit{init benign seeds}: number of benign seed nodes initially in the target’s DB. \textit{pre-restart DB}: avg. DB fill and seed occupation (\%) over 10 runs before restart. \textit{post-restart fill}: avg. fill rate of the last 2, 5, 8, and 17 buckets after restart (10 runs). \textit{outbound selection rate}: prob. that an outbound connection chooses a peer from those buckets. \textit{behavior changed?}: whether attacker behavior was altered. \textit{with/without DB poisoning}: whether step-\ding{172} was applied.)}
\label{tab:discovery_filling}
\renewcommand{\arraystretch}{1} 
\begin{threeparttable}
\resizebox{0.95\linewidth}{!}{
\begin{tabular}{ c|c| cc  c  cccc c cccc }
\toprule

 &  \textbf{init benign seed}  &  & \textbf{DB fill} & \multicolumn{1}{c}{\textbf{seed (\%)}}  & 
& \multicolumn{2}{c}{\textbf{last-2}}  
& \multicolumn{2}{c}{\textbf{last-5}}  
& \multicolumn{2}{c}{\textbf{last-8}}  
& \multicolumn{2}{c}{\textbf{last-17}}  
\\ 

\midrule

\multirow{8}{*}{\makecell{\textbf{\textit{with} DB} \\ \textbf{poisoning}} } & \multirow{4}{*}{257} 
& \multirow{10}{*}{\makecell{\textbf{pre-restart} \\ \textbf{DB}}} &  
\cellcolor{gray!10} 33\% &  \cellcolor{gray!10} 38\% &  
\multirow{10}{*}{\makecell{\textbf{post-restart } \\ \textbf{fill}}} &  
\cellcolor{gray!10}36\% & \cellcolor{gray!10}37\% & \cellcolor{gray!10}47\% & \cellcolor{gray!10}28\% & \cellcolor{gray!10}60\% & \cellcolor{gray!10}44\% & \cellcolor{gray!10}75\% & \cellcolor{gray!10}62\% \\ 

& &   & \cellcolor{gray!10} 50\% &  \cellcolor{gray!10} 58\% &   & 
\cellcolor{gray!10}53\% & \cellcolor{gray!10}55\% & \cellcolor{gray!10}60\% & \cellcolor{gray!10}34\% & \cellcolor{gray!10}69\% & \cellcolor{gray!10}47\% & \cellcolor{gray!10}81\% & \cellcolor{gray!10}65\% \\ 

& &   &  \cellcolor{gray!10} 60\% &  \cellcolor{gray!10} 68\% &   &
\cellcolor{gray!10}64\% & \cellcolor{gray!10}65\% & \cellcolor{gray!10}77\% & \cellcolor{gray!10}40\% & \cellcolor{gray!10}84\% & \cellcolor{gray!10}52\% & \cellcolor{gray!10}90\% & \cellcolor{gray!10}67\% \\ 

& &  &  \cellcolor{gray!10} 66\% &  \cellcolor{gray!10} 73\% &  &
\cellcolor{gray!10}68\% & \cellcolor{gray!10}70\% & \cellcolor{gray!10}76\% & \cellcolor{gray!10}44\% & \cellcolor{gray!10}84\% & \cellcolor{gray!10}57\% & \cellcolor{gray!10}92\% & \cellcolor{gray!10}74\% \\

\cmidrule{2-2} \cmidrule{4-5} \cmidrule{7-14}

& \multirow{4}{*}{517} &  &  \cellcolor{gray!10} 33\% &  \cellcolor{gray!10} 42\% &  &
\cellcolor{gray!10}40\% & \cellcolor{gray!10}42\% & \cellcolor{gray!10}43\% & \cellcolor{gray!10}34\% & \cellcolor{gray!10}55\% & \cellcolor{gray!10}48\% & \cellcolor{gray!10}67\% & \cellcolor{gray!10}64\% \\

& &  &  \cellcolor{gray!10} 50\% &  \cellcolor{gray!10} 56\% &  &
\cellcolor{gray!10}58\% & \cellcolor{gray!10}60\% & \cellcolor{gray!10}65\% & \cellcolor{gray!10}43\% & \cellcolor{gray!10}74\% & \cellcolor{gray!10}51\% & \cellcolor{gray!10}79\% & \cellcolor{gray!10}64\% \\ 

& &  &  \cellcolor{gray!10} 60\% &  \cellcolor{gray!10} 64\% &  &
\cellcolor{gray!10}68\% & \cellcolor{gray!10}70\% & \cellcolor{gray!10}71\% & \cellcolor{gray!10}49\% & \cellcolor{gray!10}88\% & \cellcolor{gray!10}56\% & \cellcolor{gray!10}80\% & \cellcolor{gray!10}67\% \\ 

& &  &  \cellcolor{gray!10} 66\% &  \cellcolor{gray!10} 72\% &  &
\cellcolor{gray!10}76\% & \cellcolor{gray!10}71\% & \cellcolor{gray!10}83\% & \cellcolor{gray!10}51\% & \cellcolor{gray!10}88\% & \cellcolor{gray!10}56\% & \cellcolor{gray!10}89\% & \cellcolor{gray!10}67\% \\ 

\cmidrule{1-2} \cmidrule{4-5} \cmidrule{7-14}

\multicolumn{1}{c}{\textbf{\textit{without}}} & \multicolumn{1}{c}{-} 
&  &  \cellcolor{gray!10} 0\% &  \cellcolor{gray!10} 0\%  &  & 
\cellcolor{gray!10} 0\% & \cellcolor{gray!10} 0\% & \cellcolor{gray!10} 22\% & \cellcolor{gray!10} 22\% & \cellcolor{gray!10} 41\% & \cellcolor{gray!10} 41\% & \cellcolor{gray!10} 57\% & \cellcolor{gray!10} 57\% \\ 
\midrule

\multicolumn{2}{c}{}  & \multicolumn{4}{c}{\textbf{behavior changed?}}   & Yes & No & Yes & No & Yes & No & Yes & No \\ 

\cmidrule{3-14}

\multicolumn{3}{c}{} & \multicolumn{3}{c}{\textbf{outbound selection rate}}  & \multicolumn{2}{c}{75\%} & \multicolumn{2}{c}{96.9\%} & \multicolumn{2}{c}{99.6\%} & \multicolumn{2}{c}{100\%} \\ 

\cmidrule{1-14}
\end{tabular}
}
\end{threeparttable}
\end{table*}

\subsection{Evaluating \textbf{Step-\ding{173}} (DNS List)}\label{sec:4.2}

We evaluate the feasibility and cost of DNS list poisoning attacks. We first locate the identity information of official devp2p nodes to locate the crawler nodes for the DNS list. Then, we estimate the maximum time required to fill the DNS list with nodes under our control, providing a worst-case evaluation of the attack cost. 

\smallskip
\noindent\textbf{Locate the identity information of devp2p nodes.}\label{sec:4.2.1}
Following \S\ref{sec:3.3}, we ran five Ethereum mainnet nodes on a public server and analyzed logs. From these logs we identified the devp2p crawler IP as \textit{34.243.109.22}. This crawler uses a randomized key pair and UDP port at each startup, as confirmed by repeated observations.

\begin{table}[b]
\centering
\renewcommand\arraystretch{1}
\caption{Estimated Time to Reach DNS List Completion under Different Filling Rates}
\label{tab:2}
\resizebox{0.7\linewidth}{!}{
\begin{tabular}{c|ccccc}
\toprule
\textbf{Filling rate} & \textbf{0\%} & \textbf{25\%} & \textbf{50\%} & \textbf{75\%} & \textbf{100\%} \\ 
\midrule
\textbf{Score} & \cellcolor{gray!10} 275 & \cellcolor{gray!10} 325 & \cellcolor{gray!10} 506 & \cellcolor{gray!10} 831 & \cellcolor{gray!10} 2688 \\ 
\textbf{Time} (days) & \cellcolor{gray!10} 54  & \cellcolor{gray!10} 64  & \cellcolor{gray!10} 100 & \cellcolor{gray!10} 166 & \cellcolor{gray!10} 538 \\ 
\bottomrule
\end{tabular}
}
\end{table}

\smallskip
\noindent\textbf{Estimation of DNS list filling time.}\label{sec:4.2.2}
Considering the devp2p tool management in \S\ref{sec:2.4.3}, to maximize visibility and scoring, our objective is to ensure that attacker-controlled nodes are consistently online and discoverable, receiving the maximum 3-point score during every crawler session. This allows us to estimate the minimum time required to fill the DNS list under ideal conditions.


To uphold ethical standards, we did not perform filling attacks on the official devp2p crawler.Instead we built a minimal controlled testbed on two public servers: one runs the target node (no IP restrictions, generates a random node ID and UDP port at each startup, and uses the latest DNS summary as crawled-node input), and the other hosts five attacker nodes. The attackers continuously \texttt{Ping} the target and, upon \texttt{FindNode} requests, reply with \texttt{Neighbors} messages listing all five attacker entries to maximize exposure. The experiment was repeated for 10 trials.



In 5 out of 10 experiments, at least some attacker nodes successfully entered the target's discovery table. In one experiment, all five attacker nodes reached the maximum score (3 points), demonstrating the feasibility of DNS list poisoning under controlled conditions.

We analyze the score distribution of nodes in the Sepolia testnet DNS list as of July 17, with scores ranging from 275 to 2,688. Assuming each crawl adds only one point to benign nodes and that an attacker can earn 3 points per crawl (a net gain of 2 points), the average daily score advantage is 5 ($=(30/10 - 1) * 5/2$) points. Based on this, we estimate the time required to replace 0\%, 25\%, 50\%, 75\%, and 100\% of DNS list entries, as shown in Table~\ref{tab:2}.

\subsection{Evaluating \textbf{Step-\ding{174}} (Available Slots)}\label{sec:4.3}

We analyzed the number and distribution of available incoming connection slots on both Ethereum mainnet and the Sepolia testnet, and launched a large-scale slots occupation attack targeting all reachable nodes on Sepolia. 
We probed more than 2,000 public Ethereum nodes and discovered that over 80\% limit available incoming slots to 10 or fewer, supporting the feasibility of large-scale slot occupation. (details in Appendix~\ref{appendix:available_slots_measurement}).

On July 17, 2025, we screened a total of 1,268 nodes on Sepolia testnet from the official DNS list, measured whether these nodes had available slots. We detected 321 nodes with available slots. We started 200 Sepolia testnet nodes on a public network server and configured all 1,268 nodes as static nodes.

We tried to occupy the incoming available connection slots of nodes in the whole network in real world.  After two hours, about 90\% of the nodes with available slots were occupied, leaving 34 nodes still with available slots. This means that an available slots occupation attack is feasible on Sepolia testnet.


\subsection{Evaluating \textbf{Step-\ding{175}} (Outgoing Connections)}\label{sec:4.4}

We evaluate the effectiveness of outgoing connection hijacking under varying DNS list and DB filling conditions. We conduct experiments in two conditions: before and after network-wide available slots occupation, to assess the impact on hijacking success.

\begin{packeditemize}
    \item \textit{DNS list configuration.} Since DNS list filling takes time in practice, we simulate an $n\%$ poisoning of Sepolia's 250-node DNS list by replacing the lowest-scored $250\times n\%$ entries with attacker-controlled nodes.
    \item \textit{DB filling configuration.} The target node is initialized with 250 seed nodes, and we perform two rounds of DB filling to achieve a 50\% DB poisoning rate.
\end{packeditemize}

\noindent\textbf{Before available slots occupation.}\label{sec:4.4.1}
We conduct 20 experiments on the Sepolia testnet under DNS list poisoning rates of 0\%, 25\%, 50\%, 75\%, and 100\%, with the DB filling rate fixed at 50\%. Table~\ref{tab:77} presents the results. As expected, the success rate of outgoing connection hijacking increases as the DNS list poisoning rate rises.


We also analyzed the source of benign connections in the 56 experiments where the target node's incoming connections were not fully occupied. We found that, when the DB filling rate is 50\%, 13 of these experiments (23\%) had benign connections originating from the discovery table, while 51 experiments (91\%) had them from the DNS list. 

These results indicate that a 50\% DB filling rate can effectively limit the target node’s ability to establish benign connections via the discovery table.


\smallskip
\noindent\textbf{After available slots occupation.}\label{sec:4.4.2}
We evaluated outgoing connections hijacking based on the available slots occupation in \S\ref{sec:4.3}. Experimental results are shown in Table~\ref{tab:77}. 

In the case where the DNS list filling rate is 25\% and the DB filling rate is 50\%, the success rate of outgoing connections after occupying available slots increased by 10\%; in comparison, when the DNS list filling rate is 50\% and the DB filling rate is 50\%, the success rate of outgoing connections after occupying available slots has increased by 50\%. The improvement effect is quite significant.


\begin{table}[htbp]
\centering
\caption{Success Rates of Outgoing Connection Hijacking \textit{before} and \textit{after} Available Slots Occupation (FR: filling rate)}
\label{tab:77}
\renewcommand\arraystretch{1}
\resizebox{0.7\linewidth}{!}{%
\begin{tabular}{c|ccc}
\toprule

\textbf{\makecell{DNS list FR}} & \textbf{\makecell{DB FR}} &\textbf{\makecell{\textit{beofore}}} & \textbf{\makecell{\textit{after} }} \\ 
\midrule
0\% & 50\% & \cellcolor{gray!10} 0/20 (0\%) & \cellcolor{gray!10}/ \\ 
25\% & 50\% & \cellcolor{gray!10}4/20 (20\%) & \cellcolor{gray!10}6/20 (30\%) \\ 
50\% & 50\% & \cellcolor{gray!10}8/20 (40\%) & \cellcolor{gray!10}19/20 (95\%) \\ 
75\% & 50\% & \cellcolor{gray!10}13/20 (65\%) & \cellcolor{gray!10} / \\ 
100\% & 50\% & \cellcolor{gray!10} 19/20 (95\%) & \cellcolor{gray!10}/ \\ 
\bottomrule
\end{tabular}
}
\end{table}

\subsection{Evaluating \textbf{Step-\ding{176}} (Incoming Connections)}
\label{sec:4.5}

We then evaluated attacks on both Sepolia testnet and Ethereum mainnet by monitoring if an attacker could fully occupy incoming connection slots after target nodes restarting.

As discussed in \S\ref{sec:2.3.1}, Ethereum nodes reject all incoming connection requests from the same IP address within 30 seconds. To conserve public IP resources during our experiment, we disabled this rate-limiting restriction on the target nodes. We used two public servers: Server~1 and Server~2. On Server~1, we deployed three full nodes as targets for the incoming connection hijacking test. On Server~2, we launched 40 attacker nodes configured to pre-establish peer connections by registering the target nodes as static peers using the \texttt{admin.addPeer} command in advance.

We first start the attack nodes and configure the target nodes as static nodes, and then start the target nodes. The experiment results are showed in Table~\ref{tab:11}.

\begin{packeditemize}
    \item On the Sepolia testnet, we conducted 30 trials. In all cases (100\%), attacker nodes successfully hijacked all incoming connections of the target nodes within 30 seconds of restart.
    \item On mainnet, we conducted 40 trials. In 24 of them (60\%), attackers fully occupied all incoming connection slots within two days of the target’s restart. Among these 24 successful trials, 18 (45\% of the total) achieved full occupation within 30 seconds of restart.
\end{packeditemize}

\begin{table}[htbp]
\centering
\renewcommand\arraystretch{1}
\caption{Incoming Connection Hijacking}
\label{tab:11}
\vspace{-0.1in}
\resizebox{\linewidth}{!}{%
\begin{tabular}{c|ccc}
\toprule
\textbf{Network} & \multicolumn{2}{c}{\textbf{\makecell{Success rates}}} & \textbf{\makecell{Restart interval of target nodes }} \\ 
\midrule
Sepolia  &  30/30 (100\%) &  \cellcolor{gray!10} 30/30 (100\%) &  \cellcolor{gray!10} within 30s \\ 
\cmidrule{1-3}
\multirow{3}{*}{mainnet} & \multirow{3}{*}{24/40 (60\%)} &  \cellcolor{gray!10} 18/40 (45\%) &  \cellcolor{gray!10} within 30s \\ 
& &  \cellcolor{gray!10} 23/40 (57\%) &  \cellcolor{gray!10} within 1 day \\ 
& &  \cellcolor{gray!10} 24/40 (60\%) &  \cellcolor{gray!10} within 2 days \\ 
\bottomrule
\end{tabular}
}
\end{table}

\section{Further Discussion}
\label{sec-mitigation}

\subsection{Implications of Version Update Frequency}\label{subsec-version}
The frequency of Ethereum client updates, particularly for Geth\footnote{See: \url{https://github.com/ethereum/go-ethereum/releases/}}, directly impacts how often nodes are restarted, which is an event that can facilitate eclipse attacks due to temporary loss of benign connections.
To assess this, we analyzed the release history of Geth from Ethereum’s Merge (Sep. 15, 2022) to date (May 5, 2025), spanning over 32 months. Within this period, 56 versions were released, averaging approximately 1.75 per month. This relatively high release cadence implies that mainnet nodes undergo frequent restarts, as each update generally necessitates one, temporarily severing all active peer connections.

This presents a recurring window of opportunity for attackers to exploit, as the node becomes momentarily vulnerable to peer table manipulation and connection hijacking, thereby increasing the feasibility of our proposed eclipse attack.


\subsection{IP Resources Required for Our Attack}

We estimate the public IP resources needed to perform each step of the proposed eclipse attack.

\begin{packeditemize}
\item[\ding{172}] \textit{DB pre-filling.}  
Sepolia testnet contains about 2,000 active nodes (\S\ref{sec:4.1.1}), The Sepolia testnet has a node size of approximately 2000 nodes, and attackers require 208 public IP addresses with a DB fill rate of 50\%.
If the attacker wants to achieve a 50\% DB fill rate, they need to fill about 2000 nodes in 16 rounds, each round containing 128 nodes (the discovery table allows up to 10 IP addresses from the same/24 subnet, so 208 ($=13*16$) public IP addresses are required, each IP running 10 nodes with different ports enabled for each node). For the mainnet, which includes roughly 6,000 nodes (\S\ref{sec:4.3.1}), the attacker would need 624 public IP addresses (i.e., $208 \times 3$) under the same 50\% filling rate assumption.

\item[\ding{173}] \textit{DNS list poisoning.}  
Due to the constraint that no more than 10 IPs from the same /24 subnet can appear in the discovery table, at least $272/10 = 28$ distinct public IP addresses are required.

\item[\ding{174}] \textit{Available slots occupation.}  
To fill the available incoming slots across the network, the attacker needs to control 200 active nodes. A node rejects repeated incoming connections from the same IP within 30 seconds (\S\ref{sec:2.3.1}). Due to limited IP resources (\S\ref{sec:4.3}), all attack nodes were deployed using a single public IP. However, assuming no IP constraints, 200 unique public IP addresses would allow for rapid occupation of all available slots.

\item[\ding{175}] \textit{Outgoing connections hijacking.}  
Similar to the DNS filling step, this step also requires 28 public IP addresses due to the same /24 subnet restriction ($272/10$).

\item[\ding{176}] \textit{Incoming connections hijacking.}  
Each Ethereum node allows up to 34 incoming connections by default. In our experiment, we use 40 attack nodes to ensure full occupation. To prevent 30-second rejection due to repeated IPs, each node requires a unique public IP address—thus, 40 public IPs are needed for this step.
\end{packeditemize}

Among the five steps, Step-\ding{172} and Step-\ding{174} can reuse the same set of public IP addresses. We estimate the total number of distinct public IPs required for the eclipse attack as follows :
\begin{packeditemize}
    \item \textbf{Sepolia testnet}: $208 + 28 + 28 + 40 = \mathbf{304}$ public IPs;
    \item \textbf{mainnet}: $624 + 28 + 28 + 40 = \mathbf{720}$ public IPs.
\end{packeditemize}


\section{Countermeasures}
\label{sec-mitigation}
 
\subsection{Against the Adversarial DB Filling}


\noindent\textbf{Establish a blacklist mechanism for high-frequency \texttt{pings}.}
A node records the timestamps of \texttt{ping} messages received from each peer. 
If a peer sends more than 5 \texttt{pings} within 1 minute, it is blacklisted, removed from the discovery table, and its UDP messages are dropped. 
The initial blacklist duration is 1 minute; repeated violations double the duration (2, 4, 8 minutes, etc.) up to a maximum cap (e.g., 24 hours).

\subsection{Against the DNS List Poisoning}
At present, there are no restrictions on IP resources in the DNS list, and we believe that this type of attack can be mitigated by limiting the attacker's resources.

\begin{packeditemize}
    
\item \textbf{Limit the number of nodes with the same IP address in DNS list.} This requires a large amount of IP resources for attackers to effectively carry out DNS list stuffing attacks.


\item \textbf{Establish a DNS node blacklist mechanism.} 
Ethereum can introduce a community-driven reporting system where users submit suspicious DNS nodes along with log evidence. Once the same node receives at least $N$ independent reports, it is added to a DNS blacklist. This threshold mechanism reduces false positives, while ensuring that malicious nodes are gradually excluded from the DNS list.

\end{packeditemize}


\section{Related Work}
\label{sec-rw}

\noindent\textbf{Eclipse attacks against Ethereum.}
Marcus\textit{et al.}~\cite{marcus2018low} demonstrated a low-resource eclipse on Geth v1.6.6: with only two attacker hosts they could monopolize a target’s incoming and outgoing connections by exploiting (1) the lack of a post-restart incoming-connection cap and (2) the fact that Geth’s UDP listener becomes active before bootstrapping, allowing unsolicited Ping-based connections. Dahlke \textit{et al.}~\cite{dahlke2018low} showed the method also applies to Parity. Henningsen \textit{et al.}~\cite{henningsen2019eclipsing} proposed the “False Friends” attack (Geth v1.8.0), using two IPs from different /24 subnets to occupy incoming slots without requiring a restart, and noting that Geth’s outgoing selection (one head per 17 buckets) can be abused by unsolicited Pings to seize outgoing slots.


\noindent\textbf{Eclipse attacks and more.}
Heilman \textit{et al.}~\cite{heilman2015eclipse} were the first to introduce the concept of the eclipse attack in the context of Bitcoin. Their method involves populating the target node’s peer table to monopolize its outgoing connections.
Tran \textit{et al.}~\cite{tran2021routing} proposed a more stealthy variant, termed \textit{Erebus}, which leverages the capabilities of autonomous system (AS)-level adversaries. By populating the target’s peerlist with IP addresses owned by a malicious AS and intercepting traffic at the network layer, the attacker gradually positions itself as a persistent intermediary, ultimately gaining control over all of the target’s connections.
Saad \textit{et al.}~\cite{saad2021syncattack} first introduced the idea of occupying idle connection slots across the network. 
However, they did not provide experimental validation on the Bitcoin network.
Shi \textit{et al.}~\cite{shi2025eclipse} extended eclipse attack to Monero's network. They proposed a connection reset attack, where the attacker actively forces the target node to drop existing benign connections without requiring a restart. This enables malicious peers to rapidly seize those freed slots and establish control over the target’s peer set.

\section{Conclusion}

We presented the first practical eclipse attack on Ethereum's peer-to-peer network after the Merge. Our attack enables full control over targeted execution-layer nodes' incoming and outgoing connections.  We evaluated on both  testnet (Sepolia) and mainnet. We show that an attacker with 304 public IP addresses can achieve a 95\% success rate in isolating a target node on Sepolia testnet shortly after its restart. We provided actionable countermeasures. 

We also made responsible disclosures to Ethereum. They acknowledged our reported problem and are committed to improving the P2P layer, though no final fix has been released yet. 


\bibliographystyle{unsrt}
\bibliography{bib}

\appendix


\section{The Lookup Recursive Search Process of Node Filling}\label{appendix:lookup}

The node lookup of the discovery table is an iterative process. The recursive search algorithm process is as follows:
\begin{packeditemize}
\item[i)] Randomly generate a node ID;
\item[ii)] Find the 16 nodes closest to the ID from the node list, with a set of N\textsubscript{0} nodes. Sort them in order of their distance from the node ID in i), from near to far;
\item[iii)] In N\textsubscript{0}, in the order of ii), select a node that has not been queried (and has not sent a \texttt{FindNode} message to that node), send a \texttt{FindNode} message, label the node as queried, and merge the 16 nodes in the received \texttt{Neighbors} messages with N\textsubscript{0}. Find the 16 nodes closest to the ID from the merge and replace the nodes in N\textsubscript{0}. The nodes in the received neighbor messages are stored in the discovery table of the local node;
\item[v)] Repeat the iteration of iii) until all nodes in N\textsubscript{0} have been queried, and the iteration ends.
\end{packeditemize}

\section{The Reasons why Incoming Connection Attempts fail}
\label{appendix:incom_fail}

First, if the node has already reached its incoming connection limit, new requests are automatically denied. Second, the node enforces a rate-limiting policy that rejects all incoming connections from the same IP address within a 30-second interval to prevent repeated rapid attempts. Finally, connections may be rejected due to protocol-level issues, such as mismatched network IDs or fork IDs, receipt of malformed or excessively long handshake messages (e.g., during the ETH or Snap protocol negotiation), duplicate connection attempts, or connections from the same peer ID.

Further details on these rejection conditions can be found in the handshake failure logic of the RLPX\footnote{RLPX protocol handshake failure: \url{https://github.com/ethereum/go-ethereum/blob/v1.14.3/p2p/peer_error.go\#L75}}, ETH\footnote{ETH protocol handshake failure: \url{https://github.com/ethereum/go-ethereum/blob/v1.14.3/eth/protocols/eth/protocol.go\#L67}}, and Snap\footnote{Snap protocol handshake failure section, refer to \url{https://github.com/ethereum/go-ethereum/blob/v1.14.3/eth/protocols/snap/protocol.go\#L59.}}.

\begin{figure*}[!hbt]
\centering
\begin{subfigure}[b]{0.45\textwidth}
    \centering
    \includegraphics[width=\linewidth]{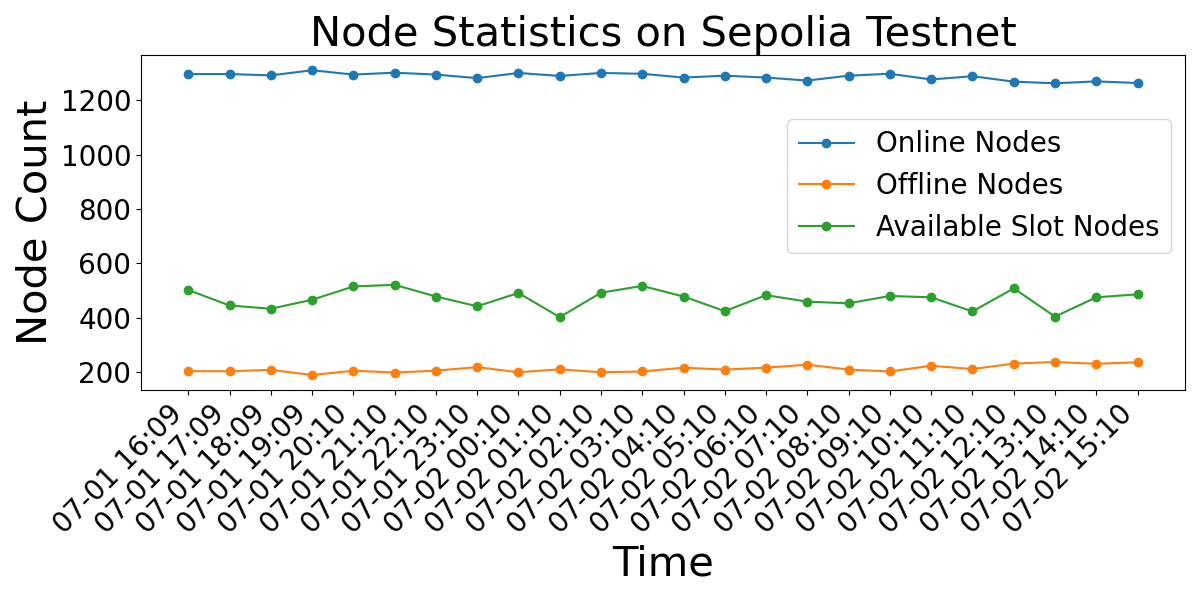}  
    \caption{Node Statistics on Sepolia Testnet}  
    \label{fig:node_workflow}
\end{subfigure}
\hfill
\begin{subfigure}[b]{0.45\textwidth}
    \centering
    \includegraphics[width=\linewidth]{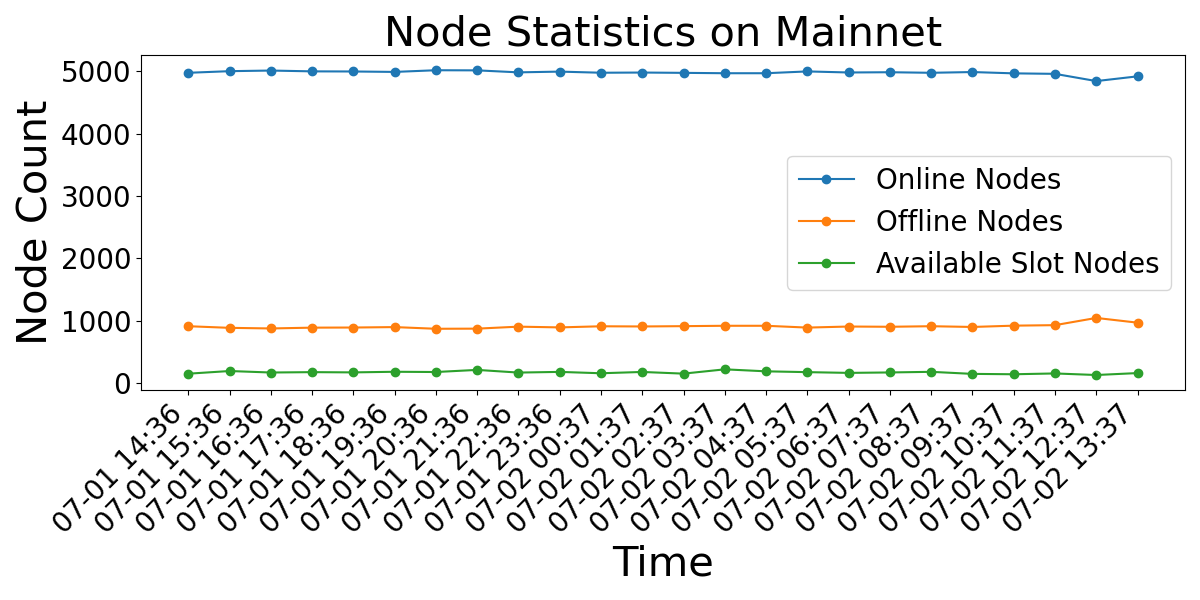}  
    \caption{Node Statistics on Mainnet}  
    \label{fig:example_figure}
\end{subfigure}
\caption{Node Statistics in Ethereum P2P Network}  
\label{fig:5}
\end{figure*}

\begin{figure*}[!hbt]
\centering
\begin{subfigure}[b]{0.45\textwidth}
    \centering
    \includegraphics[width=\linewidth]{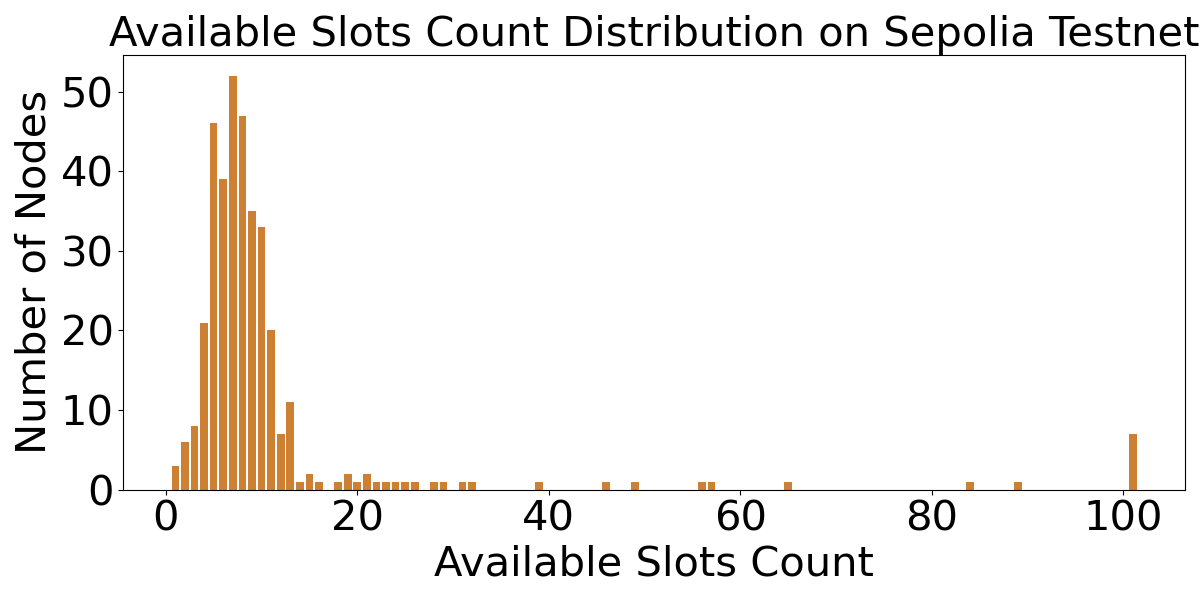}  
    \caption{Available Slots Count Distribution on Sepolia Testnet}  
    \label{fig:61}
\end{subfigure}
\hfill
\begin{subfigure}[b]{0.45\textwidth}
    \centering
    \includegraphics[width=\linewidth]{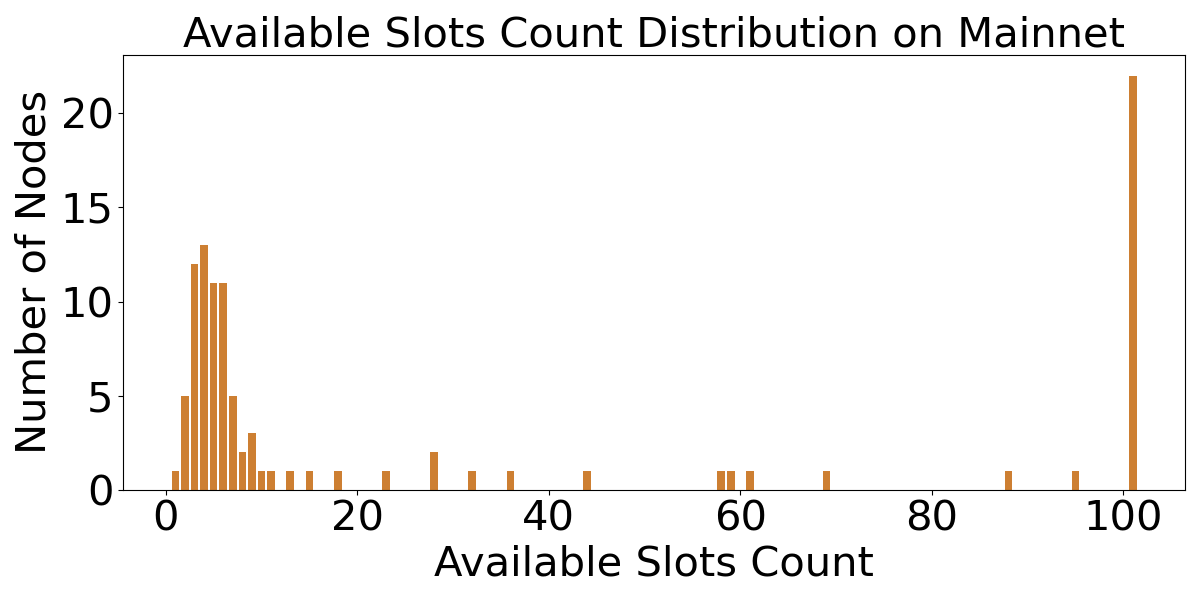}  
    \caption{Available Slots Count Distribution on Mainnet}  
    \label{fig:62}
\end{subfigure}
\caption{Available Slots Count in Ethereum P2P Network}  
\label{fig:6}
\end{figure*}

\section{Selecting Outgoing Connections}\label{appendix:select_outcon}
There are two outgoing connection selection resources for an Ethereum node, namely, lookup discovery and DNS discovery. We will introduce the two node-discovery methods below.
 
\smallskip
\noindent\textbf{Node source based on lookup discovery.}
The node source based on lookup discovery can cache up to 16 nodes. When the node source is empty, the local node will enable lookup discovery. Each lookup discovery targets the same target ID and requires multiple iterations. Below is an introduction to the iterative process of each lookup.
At the beginning of each lookup, the first step is to determine whether the local list has been searched:

\begin{packeditemize}
\item\textit{No}, then this iteration is the initial iteration round for this lookup discovery. A node ID is randomly generated as the target ID for this lookup discovery, and the 16 nodes closest to the target ID are selected from the local node list. They are sorted from small to large according to their distance from the target ID and added to the cache list result\_entries for this lookup discovery. Meanwhile, these 16 nodes are added to the \texttt{lookup\_buffer} and marked as seen. This round of iteration ends.

\item\textit{Yes}, select a node from the cache list that has not been queried in order, send it a \texttt{FindNode} message, and receive \texttt{Neighbors} messages. Combine the 16 nodes in the \texttt{Neighbors} messages with the nodes in the cache list, retain the 16 nodes closest to the target ID, and sort them from small to large according to their distance from the target ID. Meanwhile, these 16 nodes are taken from nodes that are not in the list, added to the \texttt{lookup\_buffer}, and marked as seen. This round of iteration ends.
\end{packeditemize}

The end condition for each lookup is that all nodes in the cache list searched in this lookup have been queried.
 
\smallskip
\noindent\textbf{Node source based on DNS discovery.}
After the local node is started and initialized in the node list, it will retrieve nodes from the URL of the candidate node list discovered based on DNS and store them all in the \texttt{dns\_buffer} buffer. 

The DNS discovery process is as follows:

\begin{packeditemize}
\item\textit{Obtaining DNS seeds.} In the source code of the Ethereum client, a specified DNS domain name is hard coded and used as a DNS seed to initiate the node discovery process;

\item\textit{Query DNS records and construct a node tree.} Local nodes recursively query subdomains based on branch information provided in the DNS domain name until all leaf node ENR records are obtained, and construct a DNS node tree.

\item\textit{Select a node to establish a TCP connection.} The local node extracts the node enr record from the node tree and extracts information such as node ID, IP, TCP port, etc. from the enr record. The local node attempts to establish a TCP connection with the node based on this information.
\end{packeditemize}

\section{Available Slots Measurement in Ethereum P2P Network}
\label{appendix:available_slots_measurement}

\smallskip
\noindent\textbf{Measurement of nodes with available slots.}\label{sec:4.3.1}
To measure the availability of inbound connection slots, we extracted node sets for Ethereum mainnet and the Sepolia testnet from DNS lists publicly maintained by Ethereum (snapshot taken at 09:50 UTC on July 1st, 2025). Our probing program, deployed on a cloud server in the United States, monitored these nodes continuously over a 24-hour period. Each node was scanned hourly to determine its online status and the number of available incoming connection slots.

We report the average results across 24 hourly scans in Table~\ref{tab:111}. Each scan cycle took approximately 11 minutes for the mainnet node set and 6 minutes for Sepolia. The temporal variation in the number of nodes with available slots is in Fig.~\ref{fig:5}.

We observe that the online rate of nodes on both Sepolia testnet and mainnet is consistently high, at approximately 85\%. However, there exists a stark contrast in the proportion of nodes with available inbound connection slots: around 30\% of Sepolia nodes have available slots, compared to only about 3\% on mainnet. This indicates that while overall node availability is similar, Sepolia exhibits significantly greater slot openness.

\begin{table}[htbp]
\centering
\caption{Measurement of Nodes with Available Incoming Connection Slots in the Entire Network}\label{tab:111}
\begin{threeparttable}
\resizebox{\linewidth}{!}{
\begin{tabular}{c|ccccc}
\toprule
\multirow{2}{*}{\textbf{\makecell{No.}}} & \multirow{2}{*}{\textbf{Network}} & \multicolumn{4}{c}{\textbf{Average number of nodes}} \\
\cmidrule{3-6}
 & & \textbf{Online} & \textbf{Offline} & \textbf{with available slots} \\
\midrule
1500 & Sepolia & \cellcolor{gray!10} 1288 (85.9\%) & \cellcolor{gray!10} 212 (14.1\%) &\cellcolor{gray!10}  469 (31.3\%)  \\
\cmidrule{1-4}
5887 & mainnet & \cellcolor{gray!10} 4979 (84.6\%) & \cellcolor{gray!10} 908 (15.4\%) &\cellcolor{gray!10}  168 (2.8\%)  \\
\bottomrule
\end{tabular}
}
\begin{tablenotes}
       \footnotesize
       \item[] \textbf{No.}: number of detected nodes
        \item[] \textbf{available slots}: the available incoming connection slots
\end{tablenotes}
\end{threeparttable}
\end{table}

\smallskip
\noindent\textbf{Distribution of available slots quantity.}\label{sec:4.3.2}
We launched 100 attack nodes on a public network server to actively occupy the idle connection slots of Ethereum nodes with available slots. Additionally, a dedicated detection node was deployed to verify whether all available slots had been successfully filled.


The measurement results are summarized in Table~\ref{tab:10} and Fig.~\ref{fig:6}. 
It can be seen from the table that the number of nodes with less than 10 available slots on Sepolia testnet and mainnet accounts for the majority of the number of available slot nodes, which are 80.1\% and 62.7\% respectively.

\begin{table}[htbp]
\centering
\caption{Available Slots Distribution in Ethereum Nodes}\label{tab:10}
\begin{threeparttable}
\resizebox{\linewidth}{!}{
\begin{tabular}{c|ccccc}
\toprule
\multirow{2}{*}{\textbf{\makecell{No.}}} & \multirow{2}{*}{\textbf{Network}} & \multicolumn{4}{c}{\textbf{Distribution of available slots}} \\
\cmidrule{3-6}
 & & \textbf{1-10} & \textbf{11-50} & \textbf{51-99} & \textbf{100+} \\
\midrule
362 & Sepolia & \cellcolor{gray!10} 290 (80.1\%) & \cellcolor{gray!10} 60 (16.6\%) &\cellcolor{gray!10}  5 (1.4\%) & \cellcolor{gray!10} 7 (1.9\%) \\
\cmidrule{1-3}
102 & mainnet & \cellcolor{gray!10} 64 (62.7\%) & \cellcolor{gray!10} 10 (9.8\%) &\cellcolor{gray!10}  5 (4.9\%) &\cellcolor{gray!10}  23 (22.5\%) \\
\bottomrule
\end{tabular}
}
\begin{tablenotes}
       \footnotesize
       \item[] \textbf{No.}: number of nodes with available slots
        \item[] Note: The detection duration for both networks was 7 hours.
\end{tablenotes}
\end{threeparttable}
\end{table}

\section{Pseudocode for Our Eclipse Attack}
\label{appendix:pseudo_code_of_the_eclipse_attack}

We provide implementation details of our attack (cf. Algorithm~\ref{alg:eclipse_attack}), covering the main logic and several key functions. 

In Algorithm~\ref{alg:eclipse_attack}, we give the detailed implementation process of the eclipse attack against Ethereum, including the main control logic and several key subprograms. 

In the attack initialization phase, two sub threads are started at the same time: one is to execute the \texttt{db\_pre\_filling\_attack} on the target node, polluting its local DB; the other is to execute the \texttt{dns\_list\_poisoning\_attack} on the official devp2p network crawler node of Ethereum. Next, the attacker continuously monitors whether the target node is online until it detects its restart.

After the target node restarts, the attacker immediately starts three operations:
1) Execute \texttt{occupy\_available\_slots\_attack} on all nodes in the network, and try to preempt the incoming connection slots of all active nodes, so that the target node cannot re-establish the connection with the benign nodes when it restarts;
2) Perform the \texttt{discovery\_table\_poisoning\_attack} on the target node to ensure that the attacking nodes are dominant in its restarted discovery table; 3) Start to \texttt{occupy\_in\_connections} of the target node and make it connected by attacking nodes.

Upon the above operations are executed, the attack main loop starts to run: continuously obtain the current number of incoming connections and outgoing connections (\texttt{get\_occupied\_incon} and \texttt{get\_occupied\_outcon}) of the target node, and compare them with the maximum number of allowed connections (34 and 16 respectively). As long as the connection slots are still unoccupied, the attacker will continue to launch connection attempts.

When the incoming and outgoing connections of the target node are all controlled by the attacking nodes, it means that the node is completely isolated from the honest network, and the eclipse attack is successfully completed.

\begin{algorithm}[!t]
\caption{Our \textbf{Eclipse Attack} against Ethereum}
\label{alg:eclipse_attack}

\renewcommand{\arraystretch}{1.2}
\footnotesize
\setlength{\tabcolsep}{3pt}

\BlankLine
\SetAlgoNlRelativeSize{0}
\SetKw{KwTo}{in}
\SetKw{KwRet}{return}
\SetKw{KwAnd}{and}
\KwIn{
\textit{TargetNode} = (\textit{target\_ip}, \textit{target\_port})
}

\KwData{
\textit{OccupiedInConnections} = 0,\\
\textit{OccupiedOutConnections} = 0,\\
\textit{AllActiveNodes} = nodes in Ethereum Sepolia testnet,\\
\textit{OfficialDevp2pNode} = the Ethereum official devp2p crawl node,\\
\textit{N} = the rounds of db filling attack,\\
\textit{AttackNodes} = our controlled nodes,\\
\textit{MaxInConnectionsOfTargetNode} = 34,\\
\textit{MaxOutConnectionsOfTargetNode} = 16
}

\BlankLine
 \textbf{StartThread}(\textcolor{violet}{\textbf{db\_pre\_filling\_attack}}(\textit{AttackNodes}, \textit{TargetNode}, \textit{N}))\\
 \textbf{StartThread}(\textcolor{violet}{\textbf{dns\_list\_poisoning\_attack}}(\textit{AttackNodes}, \textit{OfficialDevp2pNode}))\\

 \While{\textbf{is\_target\_online}(\textit{TargetNode})}{
    \textbf{wait()}
}

 \textbf{StartThread}(\textcolor{violet}{\textbf{occupy\_available\_slots}}(\textit{AttackNodes}, \textit{AllActiveNodes}))\\
 \textbf{wait\_for}(\textit{TargetNode\_reboot()})\\
 \textbf{StartThread}(\textcolor{violet}{\textbf{discovery\_table\_poisoning\_attack}}(\textit{AttackNodes}, \textit{TargetNode}))\\
 \textbf{StartThread}(\textcolor{violet}{\textbf{occupy\_in\_connections}}(\textit{AttackNodes}, \textit{TargetNode}))\\

 \While{\textit{OccupiedOutConnections} $<$ \textit{MaxOutConnectionsOfTargetNode} \KwAnd
\textit{OccupiedInConnections} $<$ \textit{MaxInConnectionsOfTargetNode}}{
 \textit{OccupiedOutConnections} = \textbf{get\_occupied\_outcon}(\textit{TargetNode})\\
 \textit{OccupiedInConnections} = \textbf{get\_occupied\_incon}(\textit{TargetNode})
}

 \KwRet \textit{ok}

\noindent\makebox[\linewidth]{\rule{0.9\linewidth}{0.4pt}}
\BlankLine
    
\SetKwBlock{Subroutines}{Subroutines}{}
\Subroutines{
\BlankLine
\textcolor{violet}{\textbf{discovery\_table\_poisoning\_attack}}(\textit{attack\_node\_set}, \textit{target\_node})\\
\KwData{
param1: \textit{attack\_node\_set}, param2: \textit{target\_node};
}
 \ForEach{\textit{attack\_node} \KwTo \textit{attack\_node\_set}}{
 \textit{attack\_node}.\textbf{send\_ping}(\textit{target\_node})\\
 \textit{attack\_node}.\textbf{wait\_pong}(\textit{target\_node})\\
 \textit{attack\_node}.\textbf{send\_pong}(\textit{target\_node})\\
 \textit{attack\_node}.\textbf{wait\_findnode}(\textit{target\_node})\\
 \textit{attack\_node}.\textbf{send\_neighbors}(\textit{AttackNodes})\\
 \textit{attack\_node}.\textbf{wait\_enrrequest}(\textit{target\_node})\\
 \textit{attack\_node}.\textbf{send\_enrresponse}(\textit{target\_node})
}

\BlankLine
\textcolor{violet}{\textbf{db\_pre\_filling\_attack}}(\textit{attack\_node\_set}, \textit{target\_node}, \textit{N})\\
\KwData{
param1: \textit{attack\_node\_set}, param2: \textit{target\_node}, param3: \textit{N};
}
 \For{\textit{i} = 0 \KwTo \textit{N}}{
 \textcolor{violet}{\textbf{discovery\_table\_poisoning\_attack}}(\textit{attack\_node\_set}, \textit{target\_node})
}

\BlankLine
\textcolor{violet}{\textbf{dns\_list\_poisoning\_attack}}(\textit{attack\_node\_set}, \textit{official\_node})\\
\KwData{
param1: \textit{attack\_node\_set}, param2: \textit{official\_node};
}
\While{true}{
     \textcolor{violet}{\textbf{discovery\_table\_poisoning\_attack}}(\textit{attack\_node\_set}, \textit{official\_node})
}

\BlankLine
\textcolor{violet}{\textbf{occupy\_available\_slots}}(\textit{attack\_node\_set}, \textit{node\_set})\\
\KwData{
param1: \textit{attack\_node\_set}, param2: \textit{node\_set};
}
 \ForEach{\textit{attack\_node} \KwTo \textit{attack\_node\_set}}{
 \ForEach{\textit{node} \KwTo \textit{node\_set}}{
 \textit{attack\_node}.\textbf{connect}(\textit{node})
}
}

\BlankLine
\textcolor{violet}{\textbf{occupy\_in\_connections}}(\textit{attack\_node\_set}, \textit{target\_node})\\
\KwData{
param1: \textit{attack\_node\_set}, param2: \textit{target\_node};
}
 \ForEach{\textit{attack\_node} \KwTo \textit{attack\_node\_set}}{
 \textit{attack\_node}.\textbf{connect}(\textit{target\_node})
}

\BlankLine
\textcolor{violet}{\textbf{get\_occupied\_outcon}}(\textit{attack\_node\_set})\\
\KwData{
param1: \textit{attack\_node\_set};
}
\KwRet count of outbound connections to \textit{attack\_node\_set}

\BlankLine
\textcolor{violet}{\textbf{get\_occupied\_incon}}(\textit{attack\_node\_set})\\
\KwData{
param1: \textit{attack\_node\_set};
}
\KwRet count of inbound connections from \textit{attack\_node\_set}
}
\end{algorithm}







\section{Features of Ethereum 2.0's P2P Network}

We extend our discussion in \S\ref{sec:1}. We presents key features of Ethereum 2.0’s P2P network. We focus on the validator behaviors in PoS and the importance of client diversity.

\noindent\textbf{Long-term online presence of PoS validators.} 
In PoS systems, validators are expected to maintain a persistent online presence, which is critical for ensuring the security, availability, and liveness of the network. Unlike miners in PoW systems who can participate sporadically, PoS validators are continuously involved in block proposal and finalization processes. Their availability directly impacts the network’s ability to reach consensus finality.

The validators who remain offline not only miss out on rewards but also incur economic penalties~\cite{website:pos_security}. These include being gradually penalized for inactivity and potentially being slashed for prolonged downtime or misconduct. This financial incentive structure encourages validators to keep their nodes online and reliably connected.

To further ensure network resilience, Ethereum implements an inactivity leak mechanism: when a significant portion of validators go offline, their stake influence is gradually reduced, enabling the remaining active validators to finalize the chain. This design underscores the importance of validator uptime in maintaining the health and security of PoS-based blockchains.

\noindent\textbf{Client diversity.}
In Ethereum’s current architecture, clients are divided into two categories: the Execution Layer and the Consensus Layer. The Execution Layer handles transaction execution and state updates, while the Consensus Layer is responsible for verifying consensus and finalizing blocks. Execution clients such as Geth, Nethermind, Erigon, and Besu process transactions, maintain state, and propagate blocks. In contrast, consensus clients like Prysm, Lighthouse, Teku, Nimbus, and Lodestar verify blocks, maintain the beacon chain, and generate consensus.

This modular design enhances the system’s flexibility and resilience, allowing users to freely pair execution and consensus clients according to their needs. However, the usage distribution among clients remains imbalanced~\cite{website:clinet_diversity}. Clients like Geth and Prysm still dominate market share, posing a risk of centralization. As a result, the Ethereum community places strong emphasis on client diversity, encouraging node operators to adopt a variety of implementations to improve the network’s robustness and security.

\end{document}